\documentclass[apaper,4pt]{article}
\usepackage{caption}
\usepackage{multicol}
\usepackage{graphicx}
\usepackage[left=3cm, right=3cm, bottom=3cm, top=5cm]{geometry}
\usepackage{amsmath}
\usepackage{setspace}
\usepackage{array}
\usepackage{subfigure}
\makeatletter

\newenvironment{figurehere}
{\def\@captype{figure}}
{}
\makeatother
\begin{document}
\doublespacing
\begin{center}\noindent\large \textbf{ Quark-hadron Phase Transition in Relativistic Mean-field Model }\\
{\normalsize{ Saeed Uddin\footnote {$ saeed\textunderscore jmi@yahoo.co.in$}, Waseem Bashir, Jan Shabir Ahmad\footnote{On leave of deputation from Amar Singh College, Srinagar, J \& K }, Riyaz Ahmad Bhat}}\\
\noindent\normalsize{\it Department of Physics, Jamia Millia Islamia, New Delhi -110025, India}
\end{center}
{\begin{abstract}
\noindent We have studied the quark-hadron phase transition with RMFT motivated equation of state for a strongly interacting hadronic sector and lattice motivated equation of state for weakly interacting QGP sector. The interactions in hadronic sector are dominated by the exchange of scalar and vector mesons $(\sigma^*\,\,\sigma,\,\,\omega,\,\,\rho,\,\,\phi)$ thereby allowing this phase to be modelled by the interacting baryonic, pionic and Kaonic fields. The pionic and Kaonic fields are incoorporated on equal footing to baryonic field rather than including pions and Kaons as exchange particles only. The effect of interactions on quark-hadron phase transition curve was studied using Gibbs criteria for phase equillibrium. It was found that the first order quark hadron phase transition curve ends at a critical point, whose coordinates coincide with that of the lattice gauge theory result involving lattice reweighting technique.  
\end{abstract}}

\begin{center}\noindent\textbf{\normalsize{I.\,\,INTRODUCTION}}
\end{center}
{\large Since the discovery of asymptotic freedom  \cite {gross1973} in non-abelian gauge field theories it was postulated that a phase transition from nuclear state of matter to quark matter is possible. Such transitions were further argued to happen at high temperature and /or densities where each participating nuclie would loose its existense to more fundamental degrees of freedom i.e quarks and gluons and the new state of matter  dubbed as quark gluon plasma (QGP) would be formed. Since then a considerable effrort has been put forward to create and understand the properties of this new state of matter (QGP) and the corresponding phase transition.

In order to study the phase transition in general one needs to have a complete description of a given state of matter on the basis of some underlying theory and only then one can in principle comment on the nature of phase transition by studying the variation of some parameter intrinsic to the given state as it goes through the phase transition, e.g a study of variation of order parameter in case of Landau theory of phase transition \cite {landau}, which would obtain non-zero value in one phase and vanish in other phase, and the manner in which it vanishes would decide on the nature of phase transition. 

However such a complete description for hadronic matter on the basis of quantum chromodynamics (QCD) is far from being completely understood as it falls in the regime where the strong coupling constant is too large for any peturbative expansion to remain valid. This opens up an interesting field in strong interaction physics where emphasis is on developing non-perturbative techniques from first principles e.g Lattice QCD \cite {creutz1995} or studying such non-perturbative systems using some phenomenological methods, Hadron resonance gas (HRG) and Bag models \cite {chodos1974} etc. However, each of the techniques mentioned above have some limited applicability e.g. in case of lattice-QCD one can not describe a system with non-zero baryon-chemical potential , therefore it is not possible to study the entire QCD phase diagram which involves a correct description of hadronic system with large chemical potentials as well ( even though some recent advances have made it possible to extrapolate the lattice studies to hadronic systems with finite chemical potentials but still systems with large chemical potentials are yet to be understood completely). The phenomenological thermodynamical models like HRG models and Bag models describe some of the properties of hadron and quark phases very well under some simplifying assumptions but by no means offer the complete descriptions of the these phases. The various lattice and phenomenological studies done so far have yielded different results, regarding the nature of phase transition. e.g in case of HRG based model study, one has a first order phase transition from hadronic state of matter to quark matter throughout the phase diagram, even though recently it has been claimed that after certain point the nature of phase transition changes from first to second order \cite {cpsingh2009}. Where as the lattice studies predict the existence of a critical end point for this transition, which marks the shift in the nature of phase transition from first order to second order or to smooth crossover, depending upon the quark masses used in the calculation \cite {steplatt}. Although most lattice QCD calculations indicate the the existence of critical end point (CEP) for $\mu_B$ $>$ 160 Mev \cite {steplatt}, however its exact location is  yet to be established.  

In this work we focus on the problem of quark hadron phase transition, where emphasis is on developing equation of state for strongly interacting hadronic matter very near to the phase transition region, using relativistic field theoretical description, in some approximaiton scheme, such as a mean field theoretical (MFT) description. It was found that for such strongly interacting hadronic matter the interactions are dominated by the exchange of certain mesons ($\sigma^*$,\,$\sigma$,\,$\omega$,\,$\rho$,\,$\phi$). Therefore one could in principle model the entire hadronic phase in terms of interacting baryonic, pionic and Kaonic fields, where pions and Kaons are treated on equal footing to the baryonic field rather than considering these pions and Kaons as exchange particles. This scenario will supposedly remain more and more valid as we approach more and more closer to the phase transition region, which happens to be our main region of interest for studying the quark hadron-phase transition. Quite clearly as we move away from such regions our simplifying assumption will breakdown and $\pi$ meson and K meson exchange among various baryons has to be taken into account. For QGP Sector we use an equation of state that takes into acount the perturbative interactions among quarks and is consistent with lattice data. 

The remainder of this article is organized as follows: In section II we present the formalism used in this work. In section III we show the numerical results and discussion. Finally in section IV we summarize the results and give the brief concluding discusion.}   
\vspace{3mm}
\begin{center}\noindent\textbf{\normalsize{II.\,\,THE FORMALISM}}
\end{center}
{\large We present equation of state (EOS) for the hadronic phase and the QGP phase used in this work along with the definitions}
\begin{center}\noindent\textbf{ \normalsize{A.\,\,Quark-gluon phase: Quarks u,d and s (+ gluons)}}\end{center}
{ \large To study the QGP phase with three quark flavours (u,d,s) and gluons, we use a Bag model \cite {chodos1974} equation of state, with perturbative corrections of the order of $\alpha_s$ \cite {ivanov2005} . The pressure, energy density take the form \cite {satarov}}  
\normalsize
{\begin{align*}
P(T,{{\mu}_f})= & \left(1-\frac{4}{5}{\zeta}\right) \frac{N_{g}}{6{{\pi}^2}}\int_0^{\infty}\frac{k^4dk}{\sqrt{{k^2}+{m_{g}}^2}}f_{g}(k)
+\,(1-{\zeta})\sum_{f=1}^{N_{f}}\frac{N_{c}}{3{\pi}^2}\int_0^{\infty}\frac{k^4}{\sqrt{k^2+{m_{f}}^2}}dk\\
&\times \left[f_{q,f}(k)\,+\,{\overline{f}}_{q,f}(k)\right]\,-B \tag{1}      \\\\   
\varepsilon(T,{\mu}_{f})=& \left(1-\frac{4}{5}{\zeta}\right) \frac{N_{g}}{2{\pi}^2}\int_0^{\infty}k^2 dk \sqrt{k^2+m_{g}^2}\,f_{g}(k) + (1-{\zeta})\sum_{f=1}^{N_{f}}\frac{N_{c}}{\pi^2}\int_0^{\infty} k^2 dk \sqrt{k^2 + {m_{f}}^2} \\&\times \left[f_{q,f}(k)+ {\overline{f}}_{q,f}(k)\right] + B \tag{2}
\end{align*}}
{ \large where ${\zeta}= {\alpha_s}$ is a model parameter and $1-{\zeta}$ , $1-\frac{4}{5}{\zeta}$ represent the perturbative corrections to the kinetic terms of quarks and gluons, respectively. The non-perturbative vaccum effects are taken into account by the using a Bag constant B. The $ m_g$, $m_{q}$ denote the gluon and quark masses and $f_{g}(k)$ , $f_{q}(k)$ are the bosonic and fermionic distribution functions, respectively. Here $N_{g}= 2(N_{c}^2-1)$ is the number of transverse gluons and $N_c$ is the number of quark colors ($N_{c}=3$). In case of two massless quarks (u,d) and a massive (s) quark, alongwith gluons, the above expression for the pressure and energy density reduce to:}
{\begin{align*} 
P(T,{\mu}) =& \bar{N_{g}}\frac{\pi^2 T^4}{90}+ \bar{N_{f}}\left(\frac{7}{60}\pi^2 T^4 
           + \frac{1}{2}\mu^2 T^2 + \frac{1}{4{\pi^2}}\mu^4\right)+
            \frac{1-{\zeta}}{\pi^2}\int_{m_{s}}^{\infty}dE \left(E^2 -m_{s}\right)^{\frac{3}{2}}\\& \times\left(f_{k}+{\bar{f}}_{k}\right)- B \tag{3}
\end{align*}
\begin{align*}
\varepsilon({T,{\mu}})= & 3\left(\frac{\bar{N_{g}}\pi^2 T^4}{90}\right) + 3\bar{N_{f}}\left(\frac{7}{60}\pi^2 T^4 + \frac{1}{2} \mu^2 T^2 + \frac{1}{4}\frac{{\mu}^4}{\pi^2}\right) + 3\left(\frac{1-{\zeta}}{\pi^2}\right)\int_{m_s}^{\infty} dE  E^2 \\& \times(E^2 - m_{s}^2)^{\frac{1}{2}} \left(f_{k}+\bar{f}_{k}\right) +B \tag{4}
\end{align*}}

{\large\noindent here $\bar{N_{g}} = 16(1-\frac{4}{5}{\zeta})$ is the "effective" number of the gluons and $\bar{N_{f}}= 2(1-{\zeta})$ is the "effective" number of light flavours. However it should be mentioned that throughout our discussion we use QGP phase with zero net strangeness. This assumption seems to be very reasonable in case of heavy ion collision, where the colliding nuclei do not carry any strangeness and therfore no net-strangeness is carried over to the QGP phase and the new thermally generated strange quark- antiquark pairs in the QGP phase always maintain the zero net strangeness condition.}
\begin{center}\noindent\textbf{ \normalsize{B.\,\,Hadronic phase: baryons, Kaons and pions }}\end{center}

{\large In nuclear physics where there are several types of quanta of the nuclear field, differing in quantum numbers and masses (in contrast to the electrodynamics with only one field quanta), the nucleon-nucleon (NN) potential is defined as the superposition of components with different space time transformation properties and different radii of action. In the calculation of the NN potential the range of potential is divided into three parts, the external part  $ r > 2.14fm$, the intermediate part $0.71fm <r< 2.14fm$ and the internal part $r < 0.71fm$ \cite {savaushkin}. These three regions are determined by different mesons and resonances that make up the contributions to the NN potential (from the uncertianity principle it follows that the radius of the action of the nuclear force is determined by the compton wavelength of the respective quanta). At present it can be considered as an established fact that the external and intermediate parts of the NN potential in meson theory are determined by the meson exchanges, however the internal part, which corresponds to the internucleon distance less than 0.71fm, this picture of the single boson exchange is not acceptable, as many processes make contributions to the NN forces in this region. Therefore this part of the interaction is mainly described by the phenomenological methods or by employing quark-hadron dynamics. The external region is completely determined by the one pion exchange. The pion may be coupled to the nucleon field by either pseudovector or pseudoscalar coupling. However sometimes a mixture of both coupings is used \cite {savaushkin}. To describe the intermediate range of NN  potential a proper treatment of two pion exchange contribution is required along with the exchange of other mesons. However in many NN models the attraction produced by $2\pi$ exchange is simulated by scalar-isoscalar $\sigma$ mesons with the mass around 500-600 MeV \cite {savaushkin}. Keeping in view the fact that for a possible hadron-quark phase transition, hadrons must come close to each other sufficent enough so that the quarks from each individual hadron enter the state where quarks are asymptotically free, therefore the inter-nucleon distance cannot correspond to the external part of the NN potential ($r > 2.14 fm$). Thus the one pion exchange which describes the exterior part of NN potential can be safely ignored in comparison to the scalar and vector mesons $(\sigma^*,\,\,\sigma,\,\, \omega,\,\, \rho,\,\, \phi)$ exchanges. However since the production of pions and Kaons is large at finite temperature (T), hence we need to incorporate these in the system along with the other hadrons (not as exchange particle though) with their intercations to baryons governed by the exchange of scalar and vector mesons.} 

{\large Therefore in order to describe the hadronic phase we develop a relativisic field theoretical model in which baryons, pions and Kaons are included on same footing, and are interacting via the exchange of mesons. The baryons considered are $(N,\Lambda,\Sigma,\Xi,\Delta)$ and the exchange mesons include isoscalar-scalar and vector mesons $(\sigma,\,\omega)$, isovector- vector mesons $(\rho)$ and two additional hidden strangeness mesons $(\sigma ^*, \phi) $
The lagrangian for the hadronic phase can therefore be written as,}
\normalsize{\begin{align*}
\mathcal L^{Total}=\mathcal L_B+\mathcal L_K+\mathcal L_\pi \tag{5}
\end{align*}}
{ \large Where the effective lagrangian for baryons is \cite {yangshen}:} 
{\normalsize \begin{align*}
\ \mathcal L_B=& \sum_B {{\overline\Psi}_B}[i\gamma_{\mu}\partial^{\mu} - m_B -g_{\sigma B}\sigma+
g_{\sigma^{*} B}\sigma^* -g_{\omega B}\gamma_\mu\omega^\mu-g_{\phi B}\gamma_\mu \phi^\mu -
g_{\rho B}{\gamma_\mu} {\tau_i} \rho_i^\mu] \Psi_B \\&+
\frac{1}{2}\partial_\mu\sigma \partial^\mu\sigma-\frac{1}{2} m_{\sigma}^2 \sigma^2-
\frac{1}{3} g_2 \sigma^3 - \frac{1}{4}g_3\sigma^4 -
\frac{1}{4}W_{\mu\nu}W^{\mu\nu}+\frac{1}{2}m_\omega^2\omega_\mu\omega^\mu\\&+ \frac{1}{4}c_3(\omega_\mu\omega^\mu)^2-
 \frac{1}{4}R_{i\mu\nu}R_i^{\mu\nu}+\frac{1}{2}m_\rho^2\rho_{i\mu}\rho_i^\mu+ \frac{1}{2}\partial_\mu\sigma^* \partial^\mu
\sigma^*-\frac{1}{2}m_{\sigma^*}^2 {\sigma^*}^2\\&-\frac{1}{4}S_{\mu\nu}S^{\mu\nu}+\frac{1}{2}m_\phi^2\phi_\mu\phi^\mu \tag{6}\\
\end{align*}}
\large{where} \normalsize{$S^{\mu\nu}= \partial^{\mu}\phi^{\nu}-\partial^{\nu}\phi^{\mu}$, $R^{a \mu \nu}=\partial^{\mu}\rho^{a\nu}-\partial^{\nu}\rho^{a\mu}+g_{\rho}\epsilon^{abc}\rho^{b\mu}\rho^{c\nu}$  and $W^{\mu \nu}= \partial^{\mu}\omega^{\nu}-\partial^{\nu}\omega^{\mu}$}\large{ with the sum carried over the entire baryon octet. For Kaons \cite {shaffner1996} the effective lagrangian takes the form:} 
\normalsize{
\begin{align*}
\ \mathcal L_K= &\sum_K \partial_{\mu}{ K^ \dagger} \partial^{\mu}K- m_k^2{ K^ \dagger}K- g_{\sigma K}m_K{ K ^ \dagger}K \sigma -g_{\sigma^*K}m_k{ K ^ \dagger}K\sigma^*-g_{\omega k}{ K ^ \dagger}i \overleftrightarrow{\partial_\mu} K \omega^\mu\\&-g_{\rho k}{K ^ \dagger} {\tau_K}i\overleftrightarrow{\partial_\mu}K\rho^\mu-g_{\phi K}{ K ^ \dagger}i\overleftrightarrow{\partial_\mu}K\phi^\mu+(g_{\omega K}\omega_\mu+ g_{\rho K}\tau\rho_\mu+ g_{\phi K}\phi_\mu)^2  {K^ \dagger} K \tag{7}
\end{align*}
{\large Now similarly for pions one can write }
\begin{align*}
\ \mathcal L_\pi= &\sum_\pi \partial_{\mu}{ \pi ^ \dagger} \partial^{\mu}\pi- m_\pi^2{ \pi ^ \dagger}\pi- g_{\sigma \pi}m_\pi{\pi ^ \dagger}\pi \sigma-g_{\omega \pi}{ \pi ^ \dagger}i \overleftrightarrow{\partial_\mu} \pi \omega^\mu-g_{\rho \pi}{\pi ^ \dagger} {\tau_\pi}i\overleftrightarrow{\partial_\mu}\pi\rho^\mu 
\\&+\left( g_{\omega \pi}\omega_{\mu}+ g_{\rho \pi} \tau \rho_{\mu}\right)^2 {\pi ^ \dagger} \pi     \tag{8}         
\end{align*}}
{\noindent\large One can easily see that both of these coupling schemes mentioned above fulfill the Ward identity. The resulting field equations for fermions and bosons obtained after minimising the corresponding action, $S=\int \mathcal L d^4x$, are non-linear in form and because of strong coupling constants cannot be solved using perturbative techniques and hence are to be solved using an approximation scheme e.g in mean field theoretical description. Using such an approximation, the field equations for the mesons $(\sigma,\,\omega,\,\rho,\,\sigma^*,\,\phi)$ turn out to be as follows:\\ For sigma $(\sigma)$ field;}        
{ \normalsize 
\begin{align*}
{{m_\sigma}^2}\sigma^2 +g_2 \sigma^2+ g_3 \sigma^3 =& -\sum_B g_{\sigma B}\frac{\gamma}{(2\pi)^3} \int\frac{d^3k}{\sqrt{k^2+{m_B^*}^2}} m_B^*\left(n_B +\overline n_B\right)\\&-
\sum_{b=(K,\pi)}g_{\sigma b}m_b \frac{\gamma}{({2\pi})^3}\int \frac{d^3k}{2\omega_b} \left(n_b +\overline n_b\right) \tag{9}
\end{align*}
{\large similarly for omega and rho fields one has, }
{\normalsize \begin{align*}
m_\omega^2\omega + c_3 \omega^3=& \sum_B g_{\omega B} \left[\frac{\gamma}{2\pi^3} \int d^3 k \left(n_B -\overline n_B\right)\right]+
\sum_{b=(K,\pi)} 2g_{\omega b}\left[\frac{\gamma}{(2\pi)^3}\int \frac{d^3 k}{2\omega_b} \left(E_b^+n_b + E_b^- \overline n_b\right)\right]
\\&-\sum_{b=(K,\pi)}\left(2\omega g_{\omega b}^2+ 2g_{\omega b}g_{\rho b}\rho\tau_3 + 2g_{\omega b}g_{\Phi b}\Phi \right)\times \left[\frac{\gamma}{(2\pi)^3}\int \frac{d^3 k}{2\omega_b} \left(n_b + \overline n_b\right)\right] \tag{10}
\end{align*}}}
{\normalsize \begin{align*}
 m_\rho^2 \rho= &\sum_B g_{\rho B}\tau_3 \left[\frac{\gamma}{(2\pi)^3}\int d^3 k \left(n_B - \overline n_B\right)\right]+
\sum_{b=(K,\pi)} 2 g_{\rho b}\tau_3 \left[\frac{\gamma}{(2\pi)^3}\int \frac{d^3 k}{2\omega_b}\left(E_b^+n_b + E_b^- \overline n_b\right)\right]
\\&-\sum_{b=(K,\pi)}\left(2g_{\omega_b}g_{\rho b}\omega \tau_3+ 2 g_{\rho b}^2 \tau_3^2 \rho + 2 g_{\Phi b}g_{\rho b}\tau_3 \Phi\right)
\times \left[\frac{\gamma}{(2\pi)^3}\int \frac{d^3 k}{2\omega_b}\left(n_b +\overline n_b\right)\right] \tag{11}
\end{align*}}\\
{ \large For sigmastar $(\sigma^*)$ field we obtain,}
{\normalsize \begin{align*}
 m_{\sigma*}^2 \sigma ^*=& -\sum_B g_{\sigma^* B}\left[\frac{\gamma}{(2\pi)^3}\int d^3 k \frac{m_B^*}{\sqrt{k^2 + {m_B^*}^2}}\left(n_B + \overline n_B\right)\right]-
\sum_{b=(K,\pi)} g_{\sigma ^* b} m_b \left[ \frac{\gamma}{(2\pi)^3}\int \frac{d^3 k}{2{\omega_b}}\left(n_b +\overline n_b\right)\right] \tag{12}
\end{align*}}
{\large and finally for the phi field we have,}\\
{\normalsize \begin{align*}
\ m_\Phi^2 \Phi = &\sum_B g_{\Phi B}\left[ \frac{\gamma}{(2\pi)^3}\int d^3 k \left(n_B - \overline n_B\right)\right]+
 \sum_{b=(K,\pi)} 2 g_{\Phi b}\left[ \frac{\gamma}{(2\pi)^3}\int \frac{d^3 k}{2\omega_b}\left({E_b}^+ n_b +{E_b}^- \overline n_b\right)\right]
\\&- \sum_{b=(K,\pi)} \left(2g_{\omega b}g_{\Phi b}\omega +2g_{\Phi b}g_{\rho b}\tau_3 \rho + 2\Phi g_{\Phi b}^2 \right)\times \left[ \frac{\gamma}{(2\pi)^3}\int\frac{d^3 k}{2\omega_ b}\left(n_b +\overline n_b\right)\right]   \tag{13}
\
\end {align*}}\\
\noindent\large{with the distribution function for baryons and antibaryons given by:}
\normalsize{\begin{align*}
n_B=\left[ exp\left(E_B^*-v_B /{KT}\right)+1\right]^{-1}  \tag{14}\\
\overline{n}_B=\left[ exp\left(E_B^*+v_B/{KT}\right)+1\right]^{-1} \tag{15}
\end{align*}}
\noindent\large{where the effective mass and chemical potential for baryons is,}
\normalsize{\begin{align*}
m_B^*=&m_B +g_{\sigma_B}\sigma+ g_{\sigma^* B} \sigma^*  \tag{16} \\
v_B*=&\mu_B-g_{\omega B}\omega -g_{\phi B}\phi-g_{\rho B}\tau_{3 B}\rho  \tag{17}
\end{align*}}
\large{for Kaons and pions the effective mass takes the following form,}
\normalsize{\begin{align*}
m_k^*=&\sqrt{m_k^2 +m_k(g_{\sigma k}\sigma +g_{\sigma^*k} \sigma^*)} \tag{18}\\
m_\pi^*=&\sqrt{m_\pi^2 +m_\pi g_{\sigma \pi}\sigma}  \tag{19}
\end{align*}}
\large{and the effective chemical potential:}
\normalsize{\begin{align*}
v_k=&\mu_k-g_{\omega k}\omega-g_{\phi k}\phi-g_{\rho k}\tau_3 \rho  \tag{20}\\
v_\pi=&\mu_\pi-g_{\omega \pi}\omega-g_{\rho \pi}\tau_3\rho   \tag{21}
\end{align*}}
\noindent\large{The pressure for the strongly interacting hadronic matter can be derived using the energy momentum tensor given by}

\normalsize{\begin{align*}
\ T_{\mu\nu}^{total}= &-g_{\mu\nu} \mathcal L^{total}\,\, +\,\, \left(\partial_\nu \Phi_i\right) \frac{\partial \mathcal L^{total}}{\partial \left(\partial ^\mu {\Phi_i}\right)} \tag{22}
\end{align*}}
\large{as}
\normalsize{\begin{align*}
P^{total}= \frac{1}{3}  \left\langle F\right|: T_{ii}^{total} :\left|F\right\rangle  \tag{23}
\end{align*}}
{\large Where indices (i) on $\Phi^i$ is summed over and each component corresonds to
different field.
$\Phi^i$=($\Psi$,K,$\pi$ ), where (using Eq.5)} 
\normalsize{\begin{align*}
T_{ii}^{total}= T_{ii}^{baryons}+ T_{ii}^{kaons}+ T_{ii}^{pions}  \tag{24}
\end{align*}}
{\noindent\large Thereby allowing pressure to be written as a sum of three parts}
\normalsize{\begin{align*}
P^{total}= P^{baryons} +P^{kaons}+P^{pions}   \tag{25}
\end{align*}}
\large{where the pressure due to baryons and kaons turns out to be,}\\
\normalsize{\begin{align*}
P^{baryons}=& \frac{1}{3}\sum_B \frac{1}{\pi^2}\int \frac{k^4 dk}{\sqrt{k^2 +m_B^2}}{\left(n_k + \overline n_k\right)}-
\frac{1}{2} m_\sigma^2 \sigma^2- \frac{1}{3}g_2 \sigma^3- \frac{1}{4}g_3 \sigma^4 +
\frac{1}{2} m_\omega^2 \omega^2 + \frac{1}{4} c_3 \omega^4 \\&+ \frac{1}{2} m_\rho ^2 \rho^2- \frac{1}{2}m_{\sigma^*}^2
{\sigma^*}^2 + \frac{1}{2} m_\phi^2 \phi^2  \tag{26} \\\\ 
P^{Kaons}= &\frac{1}{3}\sum_K\frac{\gamma}{(2\pi)^3}\int\frac{d^3k}{{2\omega_k}}k^2\left(n_k+{\overline n}_k\right)+\frac{1}{3}\sum_K (\left(g_{\omega_K} \omega  + g_{\phi K} \phi  + g_{\rho K} \tau_3 \rho \right)^2-g_{{\sigma^*}K} m_K\sigma^*\\&- g_{\sigma K}m_K \sigma-m_K^2) \left[\frac{\gamma}{(2\pi)^3} \int \frac{d^3k}{2 \omega_k}\left(n_k +{\overline n}_k\right)\right]-\frac{1}{3}\sum_K (2g_{\omega K}\omega + 2 g_{\omega K}\omega +2g_{\rho K} \tau_3 \rho\\& + 2 g_{\phi K}\phi)\left[\frac{\gamma}{(2 \pi )^3}\int\frac{d^3k}{2\omega_k}(E^+ n_k + E^- {\overline n}_k)\right] \tag{27}
\end{align*}\\ \\
{\large Similarly the contribution to the total pressure from pions comes out to be,}

\begin{align*}
P^{pions}=&\frac{1}{3}\sum_{\pi}\frac{\gamma}{(2\pi)^3}\int\frac{d^3k}{{2\omega_k}}k^2\left(n_k+{\overline n}_k\right)+\frac{1}{3}\sum_\pi( \left(g_{\omega\pi}\omega+g_{\rho \pi}\tau_3\rho\right)^2- g_{\sigma \pi}m_\pi\sigma- m_\pi^2)\\&\times \left[\frac{\gamma}{(2\pi)^3}\int \frac{d^3k}{2 \omega_k} (n_k + {\overline n}_k)\right]-\frac{1}{3}\sum_{\pi} \left(2g_{\omega \pi} \omega + 2g_{\rho \pi} \tau_3 \rho\right)\left[\frac{\gamma}{(2\pi)^3}\int \frac{d^3k}{2 \omega_k} (E^+ n_k + E^- {\overline n}_k)\right]  \tag{28}
\end{align*}}
{\noindent\large Now the energy density for this strongly interacting hadronic matter can be calculated as follows,}
\begin{align*}\normalsize{
\varepsilon^{total}=\left\langle F\right|: T_{00}^{total} :\left|F\right\rangle   \tag{29}}
\end{align*}
{\noindent\large which after using Eq.5, simplifies to, }
\normalsize{\begin{align*}
\varepsilon^{total}=\varepsilon^{baryons}+\varepsilon^{pions}+\varepsilon^{kaons}  \tag{30}
\end{align*}}
{\noindent\large where the energy density for baryons is found out to be,}  
\noindent\normalsize{{\begin{align*}
\varepsilon_{B}=&\sum_B \frac{1}{(\pi)^2}\int k^2 {\sqrt{k^2 + {{m_B^*}^2}}}\left(n_k +{\overline n}_k \right)dk+
\frac{1}{2}{m_\sigma}^2 \sigma^2 + \frac{1}{3} g_2 \sigma^3 +\frac{1}{4}g_3 \sigma^4 +
\frac{1}{2} {m_\omega}^2 \omega^2 +\frac{3}{4}c_3 \omega^4 \\&+\frac{1}{2}{m_\rho}^2 \rho^2+ \frac{1}{2}m_{\sigma^*}^2 
{\sigma^*}^2 +\frac{1}{2}{m_{\phi}}^2 \phi^2     \tag{31}
\end{align*}
{\large similarly for Kaons and pions one has,  }
\begin{align*}
\varepsilon_{K}=&\sum_K\frac{\gamma}{(2\pi)^3}\int\frac{d^3k}{2\omega_k}\left[(E^+)^2\,n_k+ (E^-)^2\,{\overline n}_k\right]+ \left({m_k}^2 + g_{\sigma K} m_K \sigma+ g_{{\sigma^*}K}m_K \sigma^*\right)\frac{\gamma}{(2\pi)^3}\int\frac{d^3k}{2\omega_k}\\& \times(n_k+{\overline n}_k)-\left(g_{\omega K}\omega +g_{\phi K} +g_{\rho K}\tau_3 \rho\right)^2
\frac{\gamma}{(2\pi)^3} \int \frac{d^3k}{2\omega_ k}\left(n_k + \overline n_k \right) \tag{32}
\end{align*}
\begin{align*}
\varepsilon_{P}=&\sum_{\pi}\frac{\gamma}{(2\pi)^3}\int\frac{d^3k}{2\omega_k}\left[(E^+)^2\,n_k+ (E^-)^2\,{\overline n}_k\right]+ \left( g_{\sigma \pi}m_\pi \sigma + {m_\pi}^2\right)\frac{\gamma}{(2\pi)^3}\int \frac{d^3k}{2\omega_k} \left(n_k +{\overline n}_k\right)\\&-\left(g_{\omega \pi}\omega +g_{\rho \pi}\tau_3 \rho\right)^2\frac{\gamma}{(2\pi)^3}\int \frac{d^3k}{2\omega_\pi}(n_k+{\overline n}_k)   \tag{33}
\end{align*}}}
\large\noindent{In this work we will use a successful parameter set of RMFT model, TMI (Table-I) these parameters have been determined by fitting to some ground state properties of finite nuclie, including unstable nuclie. As for meson-hyperon couplings we take the naive quark model values for the vector couplings following \cite {yangshen} (Table-V). However regarding the antibaryon couplings, there is no reliable information suitable for the high density matter and therefore antibaryon-meson coupling constants motivated by G-parity transformation will be used (Table VI). It is important to note that this simple consideration based on the G-parity transformation of meson fields is certainly an idealization as there are several effects in many body systems that can distort this picture and one is forced to use modified antibaryon-meson couplings \cite {mushtin2005}. On phenomenological level this can be taken into account by multiplying the baryon-meson couplings with a modifying factor $\eta$ which will act as a free parameter in the model and whose value can be varied from $0\,\leq\,\eta\,\leq\,1$ to take control of maximally strong antibaryon couplings to non-interacting antibaryons. However this requires some detailed study and we will not take this into account here. The coupling constants for Kaons and pions will be as in Ref. \cite {rijken} and this completes our description for the hadronic phase.}\\  
\begin{center}\noindent\textbf{\normalsize{III.\,\,RESULTS}}
\end{center}
\normalsize\large{First of all we will present some of the features of the QGP phase followed by the description of hadronic phase and associated phase transition along with other observables. In case of QGP phase, the EOS contains two undetermined parameter's B and $\zeta$ which are to be fixed uniquely for making any valuable prediction. As far as bag value B is concerned, we fix the bag value to be $B=344 MeV/fm^3$ following Ref. \cite {satarov2007}, \cite {satarovmis2007}, where it was successfully used for the fluid-dynamical calculations of the heavy-ion collisions. However in view of the fact that bag value can itself be a function of chemical potential and temperature, keeping the bag value B fixed, is an approximation. We will stick with this approximation as the more general form of $B \equiv B(\mu,T)$ requires a detailed study and will be picked up in future studies. Now to determine the model parameter $\zeta$ we first see its influence on the QGP equation of state. Fig(1) shows the variation of pressure with temperature for different values of paremeter $\zeta$. Here $\zeta=0$ corresponds to the non-interacting QGP phase. To choose among the possible values of the parameter $\zeta$, which can vary in the range [0-0.3] \cite {alford}, we show in fig(2) the scaled energy density of QGP phase for $B=344MeV/fm^3$ along with the lattice calculation results. It can be easily seen that the the best fit occurs for the $\zeta =0.2$}\\ \\                   
\begin{figurehere}
\begin{center}
\resizebox{75mm}{55mm}{\includegraphics{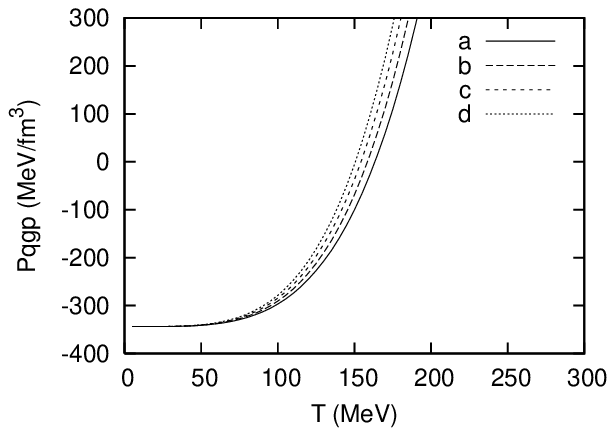}}
\end{center}
\caption{\small{Pressure in the QGP phase as function of temperature for $\zeta= 0.0,\,0.1,\,0.2,\,0.3$ for a,\,b,\,c and d  respectively. Here $\mu_q$=0 MeV.} }
\end{figurehere}

\begin{figurehere}
\begin{center}
\resizebox{75mm}{55mm}{\includegraphics{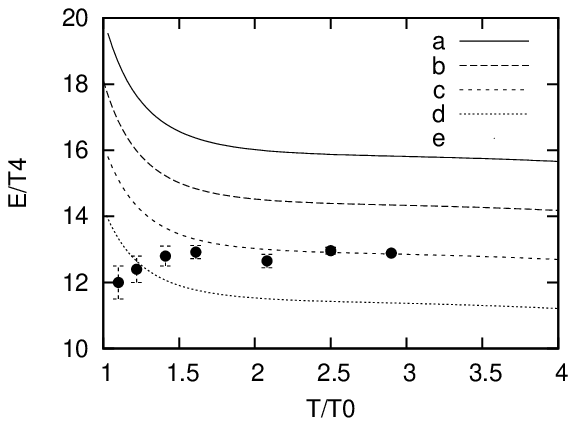}}  
\end{center}
\caption{\small{Scaled energy density for QGP phase with $\zeta=0.0,\,0.1,\,0.2,\,0.3$ for a,\,b,\,c,\,d respectively and $T_O$ is the critical temperature corresponding to each value of $\zeta$ [8]. Solid circles are for lattice data set with $N_f=3$}}
\end{figurehere}
\vspace{3mm}
{\noindent \large Throughout our calculation we will use these values of bag constant (B) and model parameter $\zeta$\footnote{In present model the parameter $\zeta$ needs to be determined by comparing scaled enery density and lattice data at finite $\mu_B$, however in abscence of lattice data at finite $\mu_B$ this value of $\zeta$ will serve as an approximate value}. Now with these fixed values of bag value (B) and model parameter $\zeta$ we next plot in Fig(3) the variation of pressure with temperature (T) for different values of chemical potential $\mu_q$.}
\begin{figurehere}
\begin{center}
\resizebox{75mm}{55mm}{\includegraphics{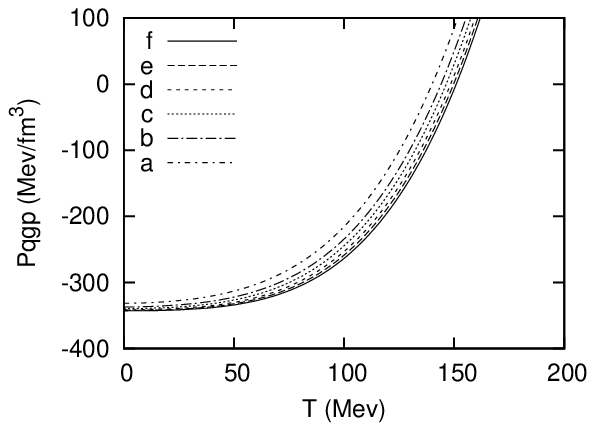}}
\end{center}
\caption {\small Pressure as function of temperature for different values of chemical potential, \,$B=344MeV/fm3$, $\zeta=0.2$,\,$\mu_q=220,190,170,150,135,120\,\, MeV$ for a,\,b,\,c,\,d,\,e and f, respectively} 
\end{figurehere}
\vspace{3mm}
{\noindent\large Next we show some of the features of the hadronic phase. The parameter set we use here are listed in table (I,\,II,\,III and IV). In Fig(4) we plot the variation of pressure in this phase.} 
\begin{figurehere}
\begin{center}
\resizebox{75mm}{55mm}{\includegraphics{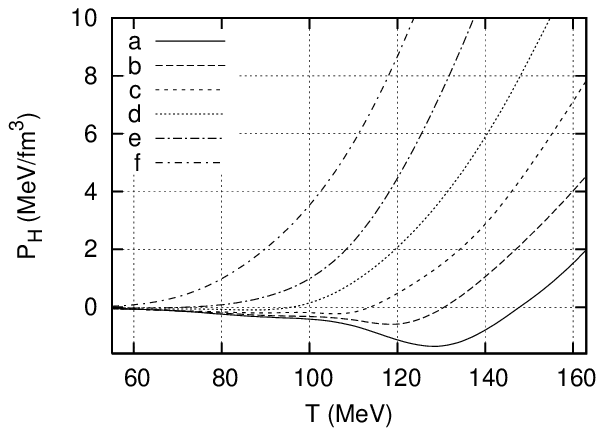}}
\end{center}
\caption {\small Pressure for hadronic phase with $\mu_q\,=\,120,135,150,170,190,220$ MeV for a,\,b,\,c,\,d,\,e and f, respectively}
\end{figurehere}
\vspace{3mm}

{\noindent \large  The $P_H$-T curves  develop the cusp which becomes more and more evident as one decreases the chemical potential $\mu_q$. This behaviour can be attributed to the interplay between attractive and repulsive attractions present in the system. Now if we calculate the slope of any general curve  with this feature as in Fig.5a we conclude that,}
{\noindent \large for the region AB the entropy desnity is:}
\normalsize{\begin{align*}
s_{AB}= S_{AB}/V= \left(\frac{\partial P}{\partial T} \right) < 0  \tag{34}
\end{align*}}
{large therefore the probability of finding the system with the pair of values ($\mu_q,\,T$), corresponding to region AB is given by:}
\normalsize{\begin{align*}
P \propto \Omega =& \exp \left(\frac{S_{AB}}{k_B}\right)     \\
                 =&\exp \left(-\frac{S}{k_B}\right)    \tag{35}
\end{align*}}
{ \noindent\large where $S_{AB}=-S$ and $k_B$ is the Boltzman's constant.
For the point B we have:}
\normalsize{\begin{align*}
s_{AB}=&0   \tag{36}
\end{align*}}
{\large which imply}
\normalsize{\begin{align*}
S= k_B\ln \Omega = 0       \tag{37}
\end{align*}}    
{\noindent\large as a result of which $\Omega=1$, where $\Omega$ are the number of states accesible to the system. Also $\Omega= E^f$, where f are the number of degrees of freedom  for the system under consideration. Therefore the point B corresponds to the configuration where the number of degrees of freedom  are very small infact zero. This configuraion can be therefore thought to be that of a system, wherein  particles  under the influence of strong attractive and repulsive forces attain the equilibrium state and arrange themselves in lattice like network. It is clear from the arguments above that the probability of finding the hadronic system with chemical potential and temperature corresponding to the region AB is very small as compared to the region BC and beyond. Therefore the effective $P_H-T$ curve can be drawn as in Fig.5b.}        

\begin{figurehere}
 \subfigure[]{\includegraphics[width=0.46\textwidth,height=0.26\textheight]{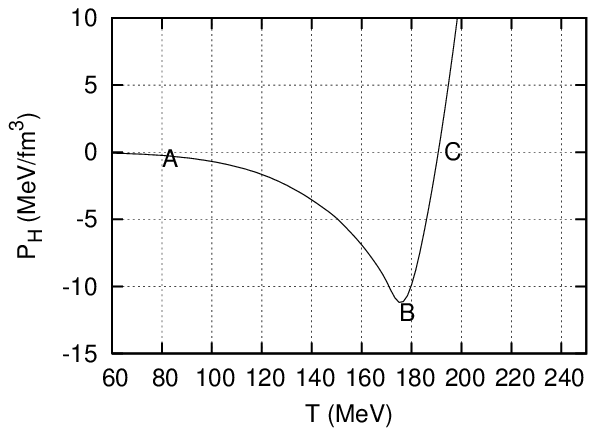}\label{}}
 \subfigure[]{\includegraphics[width=0.46\textwidth,height=0.26\textheight]{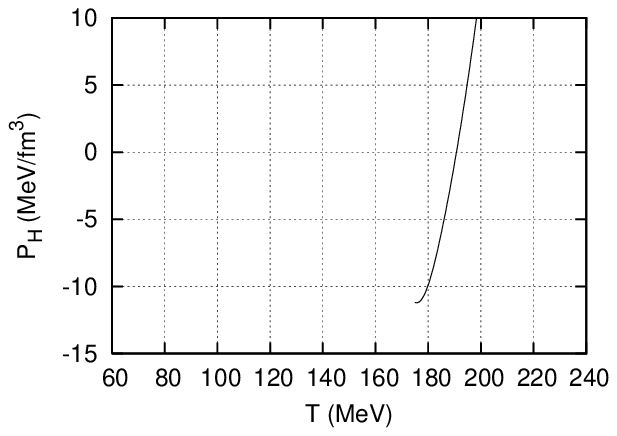}\label{}}
 \caption{$P_H$-T curve for hadronic phase for  $\mu_q$=0 MeV}\label{}
\end{figurehere}
\newpage
{\noindent\large Now to describe the quark-hadron phase transition we apply Gibbs criteria of phase transition. This hypothesis requires following set of relations to be valid at the phase coexitence points \cite{grienerbook} . }:

\normalsize{\begin{align*}
T_{Q}= &T_{H}  \\
\mu_{Q}= &\mu_{H}\\    
P_{Q}=  &P_{H}   
\end{align*}}
{\noindent\large where it is required that the state with higher pressure is more stable. Now, following three cases can arise as are shown in Fig.6. Here we show the intersection points between the P-T curves of hadronic and QGP phase. Out of the these intersection points only first two, correspond to the phase coexistence among the QGP and hadronic phase, however for intersection point in case 3, the pressure due to quarks and gluons is still less than bag pressure B, therefore the QGP phase corresponding to this intersection point is not stable and therefore cannot coexist with the hadronic phase. 
}\\

\begin{figurehere}
 \subfigure[$\mu_q$ = 135 MeV]{\includegraphics[width=0.33\textwidth,height=0.27\textheight]{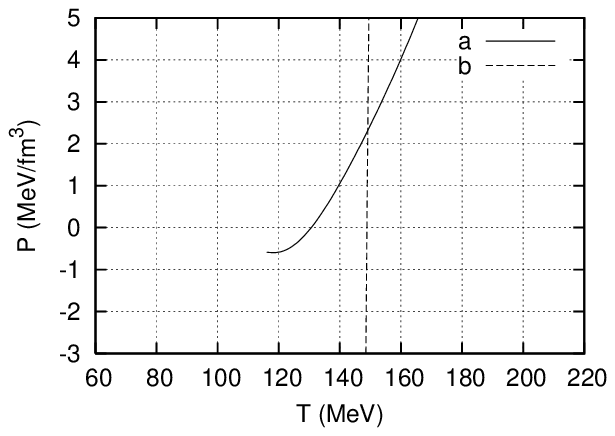}\label{first}}
 \subfigure[$\mu_q$ = 120 MeV]{\includegraphics[width=0.33\textwidth,height=0.27\textheight]{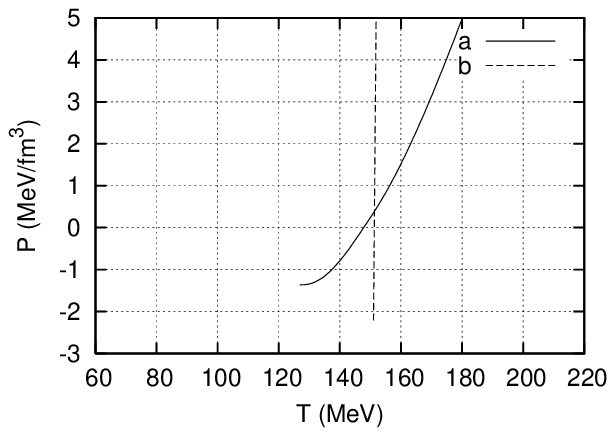}\label{second}}
\subfigure[$\mu_q$= 100 MeV]{\includegraphics[width=0.33\textwidth,height=0.27\textheight]{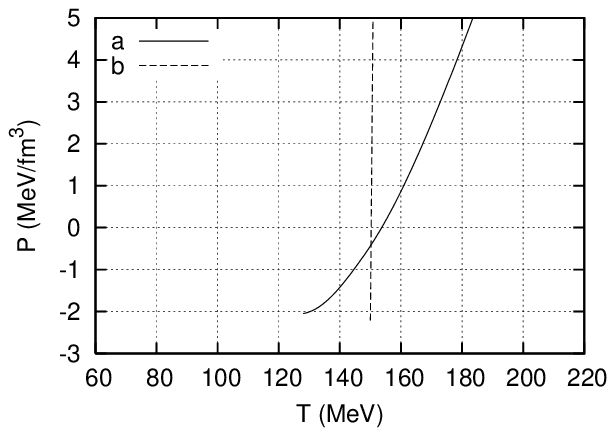}\label{second}}
 \caption{\small $P$-T curves for hadronic and QGP phases. For lower values of chemical potential the intersection point starts to appear below the P=0 axis, where the QGP phase is unstable.}\label{main_label}
\end{figurehere}
{\large }
\vspace{8mm}
{\large \noindent Now using the same procedure for the entire set of baryon-chemical potential $\mu_B$ values, the quark-hadron phase diagram turns out to be as in Fig.7 }\\\\
\begin{figurehere}
\begin{center}
\resizebox{75mm}{55mm}{\includegraphics{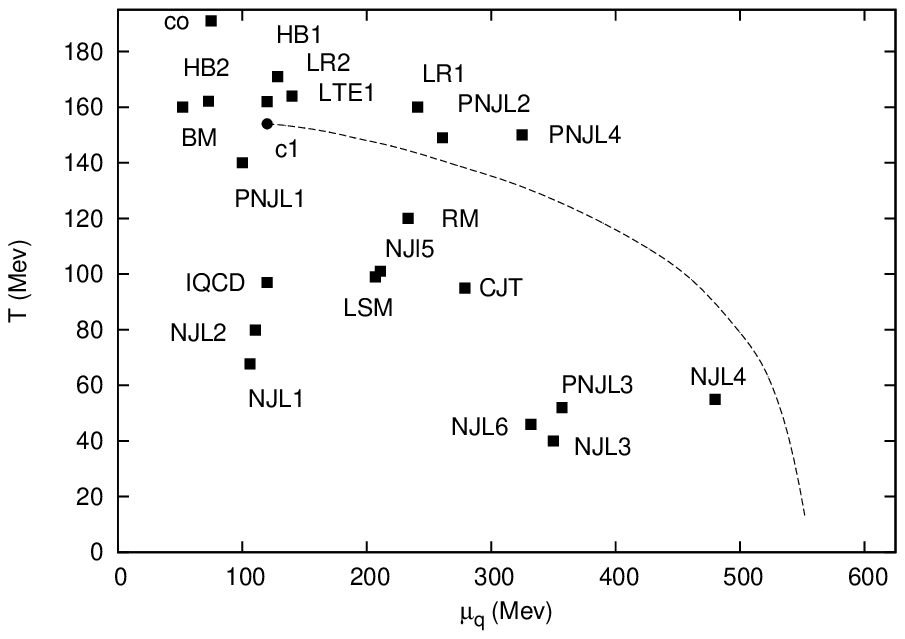}}
\end{center}
\caption{\small quark-hadron phase transition curve}
\end{figurehere}
\vspace{3mm}
{\noindent\large The first order phase transition curve ends up at the critical end point CEP ($c_1$) ($\mu_q$=120 MeV,\,T=154 MeV), which is close to the critical end point point as found using lattice Resummation technique (LR2) \cite {fodor}. In fig.7 we also show the critical end points as found in different studies \cite {srivastava}. Next we plot the variation of net baryon density ($n_B$) along this curve. Fig.8 shows the variation of $n_B$ with critical baryon chemical potential $\mu_{B_{C}}$}.    
\noindent\large{ For the sake of comparison we also plot the the variation of net baryon density $n_B$ in case of hadron resonance gas (HRG) model. It is interesting to see that the number density in case of present model saturates beyond a particular chemical potential, and this saturation happens in between the HRG curves with r=0.8\,fm and r=0.7\,fm. Therefore it can be assumed that the time independent scalar and vector potentials force the hadrons to attain an effective size. This makes sure that the number density of hadronic phase with point particles does not rise high enough to make hadronic phase stable at very high temperature or chemical potential. Next we show the variation of $n_B$ with Critical temperature T in Fig.9, here the solid dot corresponds to CEP ($c_1$).}
\begin{figurehere}
\begin{center}
\resizebox{75mm}{55mm}{\includegraphics{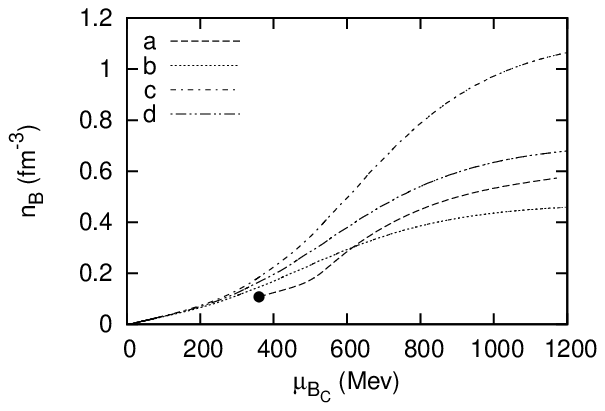}}
\end{center}
\caption{\small net baryon density v/s critical baryon chemical potential, for RMFT and HRG (with excluded volume assumption) modelled hadronic phase, where r=0.6\,\,\,fm,0.7\,fm and 0.8\,fm for c,d,b respectively}
\end{figurehere}
\begin{figurehere}
\begin{center}
\resizebox{75mm}{55mm}{\includegraphics{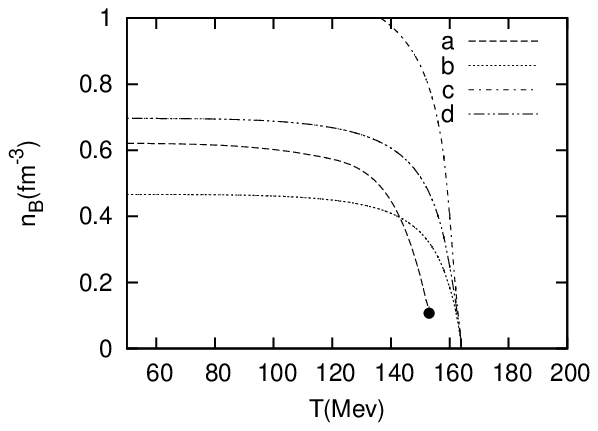}}
\end{center}
\caption{ mall net baryon density v/s critical temperature, for RMFT and HRG, here r=0.6 fm,\,0.7 fm,\,0.8fm for c,d and b respectively. }
\end{figurehere}
\vspace{3mm}
{\noindent \large To elaborate this point further, we plot the variation of proton density with temperature in our present model and compare it with HRG model for three different hard core volumes, this is shown in Fig. 10}.
\begin{figurehere}
\begin{center}
o\resizebox{75mm}{55mm}{\includegraphics{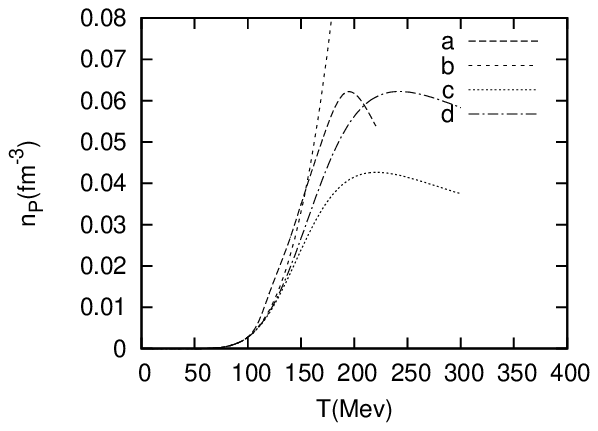}}
\end{center}
\caption{\small Proton density ($n_P$) v/s Temperature (T) for chemical potential $\mu_B\,=\,405$ Mev, r=0 fm,\,0.683 fm,\,0.8 fm for a,d and c respectively.}
\end{figurehere}
\vspace{3mm}
{\noindent \large The proton density does not rise to an arbitrary large value in our model, but saturates to a value of $0.06fm^{-3}$, this is exactly the saturation as found in HRG calculation with hard core radius r=0.683fm. Next we calculate the response of the hadronic medium to the external perturbations, for this we calculate the velocity of sound in the medium, whoose square at vanishing chemical potential is given by \cite {landaufluid}}\\
\normalsize{\begin{align*}
c_s^2= \frac{d{P}}{d{\epsilon}}       \tag{44} 
\end{align*}}
\noindent \large{since, P = P(T,\,$\mu=0$) the above equation can be written as}
\normalsize{\begin{align*}
c_s^2= \left(\frac{{{\partial{P}}/{\partial{T}}}}{{\partial{\epsilon}}/{\partial{T}}}\right)_{\mu=0}            \tag{45}
\end{align*}}
{\noindent \large In Fig.11 we plot the variation of $c_s^2$ with energy density, followed by Fig.12 where we show variation of $c_s^2$ with  temperature.}  
\begin{figurehere}
\begin{center}
\resizebox{75mm}{50mm}{\includegraphics{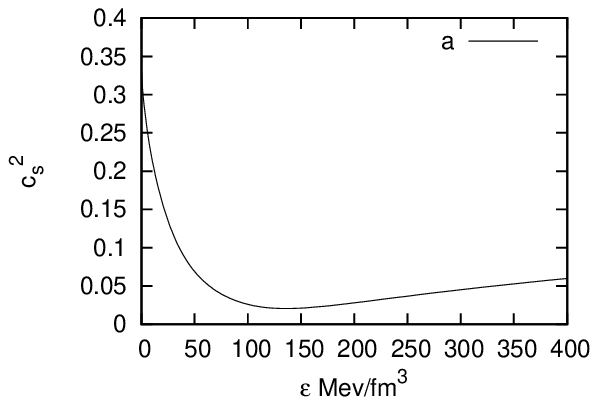}}
\end{center}
\caption{\small $c_s^2$ -$\epsilon$ for a hadronic phase with a chemical potential $\mu_q=0$ MeV }
\end{figurehere}
\vspace{3mm}
\begin{figurehere}
\begin{center}
\resizebox{75mm}{50mm}{\includegraphics{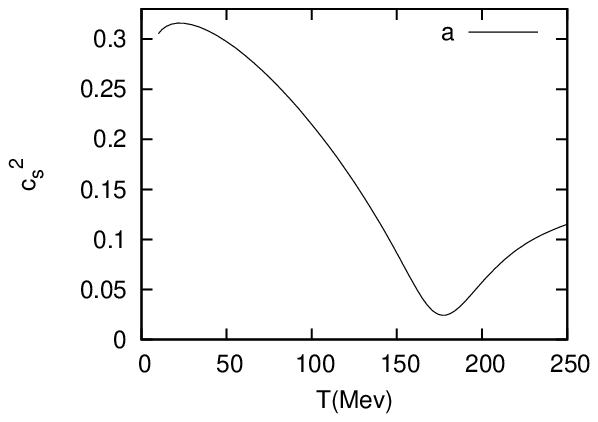}}
\end{center}
\caption{\small $c_s^2$-T for the hadronic phase with a chemical potential $\mu_q$=0 MeV }
\end{figurehere}
\noindent{}
{\large\noindent Starting with a value close to the ideal gas limit $c_s^2 \rightarrow 1/3$,there appears a sharp dip where $c_s^2$ goes to a minimum value of 0.008, beyound which it starts to rise again but not sharply though. This softening is usually attributed to the possible phase transition, however keeping into account the dynamics of present model, where scalar and vector interactions force particles to attain a state with degrees of freedom `f' equal to zero, this temperature `T' for which $c_S^2$ dips corresponds to this very configuration where we expect, given a perturbation the disturbance generated propogates through small extension of hadronic system rather being transmitted and hence a value $c_s^2=0$. It is only after this state a possible phase transition can occur.}                    
\begin{center}\noindent\textbf{\normalsize{IV.\,\, SUMMARY}}
\end{center}
{\noindent\large  We have presented a study of deconfinement phase transition from hadronic matter to the quark gluon plasma (QGP) that could be formed in the heavy ion collision. We modelled the entire hadronic phase with pions,kaons and baryons, with interactions among various hadrons carried by some specified vector and scalar mesons only. The QGP phase with perturbative intercations, was modelled in terms of MIT bag model with an equation of state consistent with the lattice data.
In such a hadronic phase we found that the interactions among various hadrons force each hadron to pick up an effective volume, the corresponding radius is found out to be r=0.683fm. This has a serious consequence as it makes sure that the number density of hadrons does not rise to an arbitray large values, which makes the hadronic phase stable again at very large temperature or chemical potential. We further found that a first order phase transition is not possible throughout the phase diagram and ends up at a critical end point (CEP) $c_1$. This is found to lie very close to the CEP as fouund in LR2[18]. However it should be mentioned that at higher temperature and/or chemical potential one needs to take into account the fluctuations about the mean value also into account, which calls for an approach beyound the mean field level, this surely needs a systematic study and is currently under investigation.}\\     
\begin{center}\noindent\textbf{\normalsize{ACKNOWLEDGMENTS}}
\end{center}
{\large Waseem Bashir is thankful to University Grants Commission for providing Project Fellowship, Saeed Uddin is thankful to the University Grants Commission (UGC), New Delhi, for the Major Research Project grant. Jan Shabir Ahmad is greatful to University Grants Commission, New Delhi, for the financial assistance during the period of deputation. Riyaz Ahmad Bhat is thankful to Council for Sceintific and Industrial Research (CSIR), New Delhi, for awarding Junior Research Fellowship.       }
 
\newpage
\begin{center}
Table I: The parameter set TMI \cite {yangshen} used in the calculation. The masses are given in MeV\\
\large
\vspace{2mm}
\begin{tabular}{|r|r|r|r|r|r|r|r|r|r|}
\hline
$m_N$ & $m_\sigma$ & $m_\omega$ & $m_\rho$ & $g_{\sigma N}$ & $g_{\omega N}$ & $g_{\rho N}$ & $g_2 (fm^{-1})$& $g_3$ & $C_3$\\
\hline
938.0&511.198&783.0&77.0&10.029&12.614&4.632 &-7.2323& 0.618 &71.308\\
\hline
\end{tabular}\\
\end{center}
\vspace{2mm}
\begin{center}
Table II: Hyperon coupling constants\\
\vspace{2mm}
\begin{tabular}{|r|r|r|r|r|}
\hline
$g_{\sigma \pi}$ & $g_{\sigma \Sigma}$ & $g_{\sigma \Xi}$ & $g_{{\sigma^*} \Lambda}$ & $g_{{\sigma^*} \Xi}$  \\
\hline
6.170& 4.472 & 3.202 & 7.018 & 12.600\\
\hline
\end{tabular}\\
\end{center}
\vspace{2mm}
\begin{center}
Table III: Kaon coupling constants:\\
\vspace{2mm}
\begin{tabular}{|r|r|r|r|r|}
\hline
$g_{\sigma K}$ & $g_{\omega K}$ & $g_{\rho K}$ & $g_{{\sigma^*} K}$ & $g_{{\phi} K}$  \\
\hline
1.93& 3.02 & 3.02 & 2.65 & 4.27\\
\hline
\end{tabular}
\end{center}
\vspace{2mm}
\begin{center}
Table IV
\vspace{2mm}
Pion coupling constants:\\
\begin{tabular}{|r|r|r|r|r|}
\hline
$g_{\pi \rho}$ & $g_{\pi \omega}$ & $g_{\pi \sigma}$ & $g_{\pi \sigma^*}$ & $g_{\pi \phi}$  \\
\hline
0.506 & -0.001& -0.170 & 0.0& 0.0  \\
\hline
\end{tabular}
\end{center}

\vspace{2mm}
\begin{center}
Table V: Vector Couplings used in the present calculations \cite {yangshen}\\

\large
\vspace{2mm}
$\frac{1}{3}g_{\omega N}$\,\,=\,\,$\frac{1}{2}g_{\omega \Lambda}$\,\,=\,\,$\frac{1}{2}g_{\omega \sigma}$\,\,=\,\,$g_{\omega \Xi}$\\
\vspace{2mm}
$g_{\rho N}$\,\,=\,\,$\frac{1}{2}g_{\rho \sigma}$\,\,=\,\,$g_{\rho \Xi}$,\,\,\,\,\,\,$g_{\rho \Lambda}=0.$\\
\vspace{2mm}
$2g_{\phi \Lambda}$\,\,=\,\,$2 g_{\phi \sigma}$\,\,=\,\,$2 g_{\phi \Xi}$= -$\frac{2\sqrt2}{3} g_{\omega N}$,\,\,\,\,\,$ g_{\phi N}$\,\,=\,\,0.\\
\end{center}

\vspace{2mm}
\begin{center}
Table VI: G-parity motivated antibaryon coupling constants \cite {mushtin2005}\\

\large
\vspace{2mm}
$g_{\sigma \overline B}$\,\,=\,\,$g_{\sigma B}$,\,\,\,\,$g_{\omega \overline B}$\,\,=\,\,-$g_{\omega B}$,\,\,\,\, $ g_{\rho 
\overline B}$\,\,= \,\,$g_{\rho B}$.
\vspace{2mm}

$2 g_{\phi \Lambda}$\,\,=\,\,$2 g_{\phi \Sigma}$\,\,=\,\,\,$g_{\phi \Xi}$\,\,=\,\,\,$\frac{-2\sqrt 2}{3} g_{\omega N}$
 $g_{\omega \overline N}$\,\,=\,\,-$g_{\omega N}$\\
\vspace{2mm}
 $g_{\phi \overline B}$\,\,=\,\,-$g_{\phi B}$\\

\end{center}
\end{document}